\documentclass[9pt,twocolumn,twoside]{pnas-new}

\templatetype{pnasresearcharticle} 

\usepackage{graphicx}	
\usepackage{amsmath}	
\usepackage{bm}
\usepackage{mathrsfs}

\graphicspath{}

\begin{document}

\title{Refractive lensing of scintillating FRBs by sub-parsec cloudlets in the multi-phase CGM}

\author[a,b,c]{Dylan L. Jow}
\author[a]{Xiaohan Wu}
\author[a,b,c,d,e,f]{Ue-Li Pen}

\affil[a]{Canadian Institute for Theoretical Astrophysics, University of Toronto, 60 St. George Street, Toronto, ON M5S 3H8, Canada}
\affil[b]{Department of Physics, University of Toronto, 60 St. George Street, Toronto, ON M5S 1A7, Canada}
\affil[c]{Dunlap Institute for Astronomy \& Astrophysics, University of Toronto, AB 120-50 St. George Street, Toronto, ON M5S 3H4, Canada}
\affil[d]{Institute of Astronomy and Astrophysics, Academia Sinica, Astronomy-Mathematics Building, No. 1, Section 4,
Roosevelt Road, Taipei 10617, Taiwan}
\affil[e]{Perimeter Institute for Theoretical Physics, 31 Caroline St. North, Waterloo, ON, Canada N2L 2Y5}
\affil[f]{Canadian Institute for Advanced Research, CIFAR program in Gravitation and Cosmology}

\leadauthor{Lead author last name}

\significancestatement{In this work, we theoretically demonstrate a novel method by which fast radio bursts (FRBs) that are scintillated by scattering in the Milky Way interstellar medium can be used to constrain physical properties of the diffuse circumgalactic medium (CGM) surrounding other galaxies. Essentially, we explicated the ways in which different models of the CGM predict scattering that can modify the amplitude of observed scintillation. As a larger population of scintillating FRBs are detected, this method will shed light on the physical state of the CGM, which contains a substantial fraction of all the baryons in the universe.}


\correspondingauthor{\textsuperscript{1}To whom correspondence should be addressed. E-mail: djow@physics.utoronto.ca}


\begin{abstract}
We consider the refractive lensing effects of ionized cool ($T \sim 10^4\,{\rm K}$) gas cloudlets in the circumgalactic medium (CGM) of galaxies. In particular, we discuss the combined effects of lensing from these cloudlets and scintillation from plasma screens in the Milky Way interstellar medium (ISM). We show that, if the CGM comprises a mist of sub-parsec cloudlets with column densities of order $10^{17}\,{\rm cm}^{-2}$ (as predicted by McCourt et al. 2018), then FRBs whose sightlines pass within a virial radius of a CGM halo will may be lensed into tens of refractive images with a $\sim 10\,{\rm ms}$ scattering timescale. When these images are formed, they will be resolved by scintillating screens in the Milky Way ISM, and will suppress the observed scintillation. We illustrate this effect in refractive lensing and argue that positive detections of FRB scintillation may constrain the properties of these cool-gas cloudlets, with current scintillation observation weakly disfavouring the cloudlet model. We propose that sheet-like geometries for the cool gas in the CGM can reconcile quasar absorption measurements (from which we infer the presence of the cool gas with structure on sub-parsec scales) and the unexpected lack of lensing signals from this gas thus far observed. 
\end{abstract}

\dates{This manuscript was compiled on \today}
\doi{\url{www.pnas.org/cgi/doi/10.1073/pnas.XXXXXXXXXX}}

\maketitle
\ifthenelse{\boolean{shortarticle}}{\ifthenelse{\boolean{singlecolumn}}{\abscontentformatted}{\abscontent}}{}

\firstpage{2}


\section{Introduction}
\label{sec:intro}

Despite containing a substantial fraction of the total number of baryons in the Universe, the physics of the multiphase gas in the circumgalactic medium (CGM) of galaxies is poorly understood. Contrary to previous expectations drawn from simple physical models, absorption line studies of the CGM on large scales regularly detect small amounts of cool gas ($T\sim 10^4\,{\rm K}$) with a small total volume, but a large covering fraction \citep[see e.g.][]{2009ApJ...690.1558P, 2013ApJ...763..148S, 2016ApJS..226...25L}. Fluorescent emission profiles rule out the possibility of the cool gas being distributed in, e.g., a thin shell surrounding the galaxy \citep{2014Natur.506...63C, 2015Sci...348..779H}. A natural conclusion from these observations is that galactic halos are filled with tiny but dense structures of cool gas. One potential physical process by which this could occur is the shattering picture proposed by Ref.~\cite{McCourt2018}, in which clouds of pressure-confined, hot gas rapidly fragment as they cool into smaller cloudlets with a characteristic length scale of $\sim c_s t_{\rm cool}$, where $c_s$ is the sound speed for the cool gas and $t_{\rm cool}$ is the cooling time. As Ref.~\cite{McCourt2018} argue, this universal length scale will lead, in the case of the CGM, to the formation of many tiny cloudlets with a characteristic radius of $r_c \sim 0.1\,{\rm pc}$ and characteristic column density $N_H \sim 10^{17} \, {\rm cm}^{-2}$. This mist formed of sub-parsec cloudlets of cool gas suffusing the CGM can explain the ubiquitous detection of a low volume of cool gas with a high covering fraction. 

Ref.~\cite{VP2019} note that if the CGM is indeed filled with sub-parsec scale cloudlets with $N_H \sim\,10^{17}\,{\rm cm}^{-2}$, then, assuming some fraction of the gas is ionized, there may be observable scattering of radio sources such as fast radio bursts (FRBs). Assuming a uniform density profile of ionized plasma within the cloudlets, Ref.~\cite{VP2019} show that cloudlets with size $r_c \sim\,1\,{\rm pc}$ may produce diffractive scattering of FRBs with a scattering timescale of $\sim 1\,{\rm ms}$, which is broadly consistent with observed FRB scattering timescales. However, observations of FRB scintillation effectively rule out the possibility that the ubiquitously observed scattering tails in FRBs can be due to scattering in the CGM of intervening galaxies \citep{Masui2015, Ocker2022, Sammons2023}. Essentially, a scattering screen somewhere along the line of sight of the FRB that produces a $\sim {\rm ms}$ scattering timescale will be resolved by the scattering screens responsible for scintillation located within the Milky Way. It is assumed that when the scintillation screen resolves the scattering screen, scintillation is suppressed. Thus, a positive detection of FRB scintillation allows for constraints to be placed on how far away the scattering screen can be from the FRB. In the case of FRB 110523, the first FRB to be confirmed to scintillate due to scattering in the Milky Way ISM, the scattering screen responsible for the millisecond scattering tail is constrained to be within $44\,{\rm kpc}$ of the FRB source, and, therefore, cannot be associated with any galaxy's CGM \citep{Masui2015}. Since a substantial number of FRBs have been confirmed to scintillate, it is unlikely that the observed scattering tails come from diffractive scattering in the CGM. Moreover, Ref.~\cite{McCourt2018} predict, using basic physical arguments and hyrodynamic simulations, the characteristic scale of the cool gas cloudlets to be $r_c \sim 0.1\,{\rm pc}$. In principle, the diffractive scattering picture employed by Ref.~\cite{VP2019} also rules out scattering by clouds of size $0.1\,{\rm pc}$, as the scattering time induced by clouds of this rise rises above $0.1\,{\rm s}$, which is inconsistent with observations. However, as the diffractive scale (the length scale below which phase variations in the scattered wavefront lead to diffractive scattering) is many orders of magnitude smaller than the size of the cloudlets \citep[$r_{\rm diff} \sim 10^{11}\,{\rm cm}$,][]{VP2019}, diffractive scattering will be highly dependent on the internal structure of the ionized plasma within the cloudlets. If, indeed, the ionized plasma is not effectively a Kolmogorov turbulent medium with outer-scale $r_c$, as proposed by Ref.~\cite{VP2019}, then the observable signature of diffractive scattering may be made arbitrarily different depending on the model of the cloudlets' internal structure. 

Thus, despite the fact that sub-parsec cloudlets of cool gas in the CGM, should they exist, will almost certainly scatter FRBs, the absence of any confirmed detection of scattering associated with the CGM has thus far not been taken as strong evidence for the lack of these cloudlets. In this work, we will discuss the refractive scattering effects of sub-parsec, cool-gas cloudlets. The refractive scale for scattering by these cloudlets is effectively set by the size of the cloudlets themselves, $r_c$, and, therefore, refractive scattering is not sensitive to the details of the internal structure of the cloudlets. We will show that even if refractive scattering leads to relatively small scattering times, e.g. down to $10\,\mu{\rm s}$, -- which would be challenging to disentangle from the observed ms scattering of FRBs from their local environment -- nevertheless, refractive scattering in the CGM will lead to a suppression of FRB scintillation. While scintillation has been used by several authors \citep[e.g.][]{Masui2015, Ocker2022, Sammons2023} to argue that the ms scattering tails observed in FRBs must be associated with the host environment, we will go further and argue that there can be no significant scattering of the FRBs along the line of sight for scintillating FRBs. In particular, whereas it is typically assumed that if the scintillation screen in the ISM resolves the primary scattering screen then scintillation is suppressed, we will quantitatively investigate this effect to clarify the magnitude of the suppression, and under what precise conditions this suppression occurs. We will use this to argue that the ubiquitous existence of cool-gas cloudlets of size $r_c \sim 0.1\,{\rm pc}$ and electron column density $N_e \sim 10^{17}\,{\rm cm}^{-2}$ is disfavoured by current observation evidence. Note, however, that since the amount of published FRB scintillation data is small (e.g. Table~\ref{tab:FRBs} shows all of the currently published FRB modulation indices), this argument is based on extrapolation from the small number of data points available. A robust conclusion can only come from further observational studies of FRB scintillation.

The paper is organized as follows: in Section~\ref{sec:lensing}, we will quantitatively study the suppression of scintillation when the CGM scattering screen is resolved by the scintillation screen using a multi-plane lensing formalism. In Section~\ref{sec:CGM}, we will use the phenomenon described in the previous section to place constraints on sub-parsec, cool-gas cloudlets in the CGM. While we argue that there cannot be strong scattering of FRBs due to the CGM, as this would lead to the total suppression of FRB scintillation, we note that FRB scintillation does appear to be relatively weaker than pulsar scintillation. That is, the few published modulation indices for scintillating FRBs appear to be in the range $0.1 < m < 0.9$ (see Table~\ref{tab:FRBs}), indicating that FRBs are generally weakly scintillated, as opposed to pulsars, which have been observed to undergo strong scintillation. Since presumably the same structures in the ISM are responsible for both pulsar and FRB scintillation, it remains to be explained why FRBs appear to scintillate more weakly than pulsars. While many effects may lead to this modest suppression of scintillation in FRBs (for example, nano-shot structures in the intrinsic burst profile of FRBs may lead to lower modulation indices, as has been observed for giant pulses in the Crab pulsar \citep{2023ApJ...945..115L}), it remains a possibility that scattering in the CGM produces a small number of refractive images, leading to a modest suppression of the modulation index. We further discuss the implications of this in Section~\ref{sec:discussion}.

\section{Scintillation in the presence of a resolving lens}
\label{sec:lensing}

The basic effect which we wish to describe is the suppression of scintillation in the ISM due to the presence of an extra-galactic lens which resolves the scintillation screen. First, we will clarify a few key terms. ``Scintillation" refers to the presence of coherent modulation of the observed intensity of a source as a function of frequency, $I(f)$. The modulation index, $m$, which we will define quantitatively later, characterizes the size of the intensity fluctuations, with $m = 1$ indicating strong scintillation. Scintillation of coherent radio sources such as pulsars and FRBs is known to be due to multi-path propagation through plasma structures in the ISM. Thus, scintillation is a \textit{lensing} effect; turbulent plasma structures in the ISM lens the source so that the observed flux is the result of an interference pattern between many images. The term ``scattering" is also used to describe multi-path propagation and is equivalent to ``lensing" in this context. Note, however, that a source may be strongly scattered without necessarily scintillating. For example, if the scattering time scale (i.e. the typical difference between arrival times of scattered/lensed images) is larger than the burst duration of the source, there will be no overlap between the images upon arrival at the observer, which is a requirement for the interference that leads to scintillation. Instead, the images will appear as a scattering tail in the time domain. Thus, while FRBs are ubiquitously observed to have exponential scattering tails, not all FRBs are observed to scintillate. 

\begin{figure*}
    \centering
    \includegraphics[width=1.6\columnwidth]{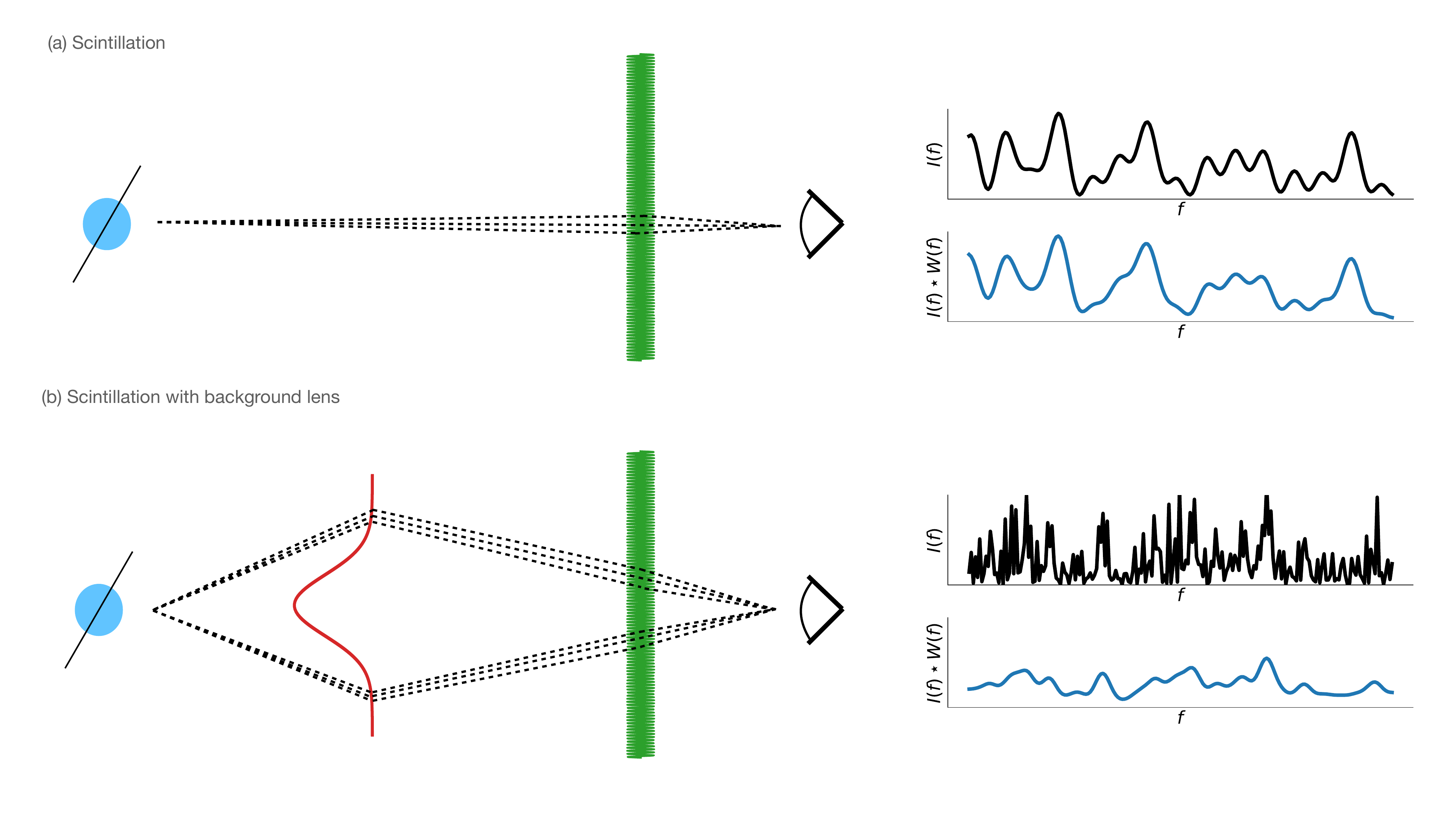}
    \caption{Diagram of the two-screen lensing. (a) When only the scintillation screen is present, the multi-path propagation of light leads to intensity modulations as a function of frequency, $I(f)$. The black intensity curves show the intrinsic intensity modulation, whereas the blue curve shows the observed intensity modulation taking into account finite frequency resolution (the observed intensity is the intrinsic intensity smoothed over the frequency resolution). If the characteristic scale of these modulations (the ``scintillation bandwidth") is smaller than the frequency resolution, then the observed intensity is effectively the same as the intrinsic intensity. (b) In the presence of a background resolving lens, an addition timescale is present, due to the multi-path propagation through the lens, as well as the scintillation screen. This leads to intensity modulations on a smaller frequency scale, which, if below the frequency resolution, results in the observed intensity modulations being suppressed relative to the case where the background lens is absent.}
    \label{fig:lenscint_diag}
\end{figure*}

Fig.~\ref{fig:lenscint_diag} shows a basic sketch of the effect we describe in the following sections. In the absence of an intervening lens, scintillation arises due to multi-path propagation through a nearby scintillation screen. However, in the presence of a background resolving lens which itself causes the source to be multiply imaged, each of these images is scintillated independently. If the scattering time scale associated with the background lens is larger than the scattering time scale associated with the scintillation screen, the result is that the intensity spectrum is modulated on a frequency-scale that is much smaller in the case of just the scintillation screen alone. If the characteristic frequency scale is smaller than the frequency resolution of the observation, then the intensity spectrum will appear less strongly modulated due to the presence of the additional lens. 

\subsection{Scintillation as a Gaussian process}
\label{sec:simple}

In Section~\ref{sec:realistic}, we introduce a simple double lens model in order to demonstrate how an additional lens can, under certain circumstances, suppress scintillation. However, we would first like to give a basic argument to clearly illustrate how this effect arises and why it is generic. 

For a strongly scintillated source, the observed voltage can effectively be described by a set of complex, normally distributed numbers, $V_i$, where $i$ indexes the frequency bin \citep{Salpeter1967, Johnson2012}. The observed intensity is 
\begin{equation}
    I_i = |V_i|^2 = {\rm Re}[V_i]^2 + {\rm Im}[V_i]^2,
\end{equation}
where the distribution of $I_i / \sigma^2$ is $\chi^2(2)$, since ${\rm Re}[V_i], {\rm Im}[V_i] \sim \mathcal{N}(0,\sigma^2)$.

Defining $\delta T \equiv (I_i - \overline{I}) / \overline{I}$, the modulation index is defined by
\begin{equation}
    m \equiv \langle \delta T^2 \rangle = \frac{{\rm Var}[I_i]}{\overline{I_i}^2} = \frac{{\rm Var}[I_i / \sigma^2]}{\overline{I_i / \sigma}^2} = 1,
\end{equation}
where the last equality follows from the fact that $I_i / \sigma^2 \sim \chi^2(2)$ and $\chi^2(k)$ distributed variables have a variance of $2 k$ and mean of $k$. Thus, for a strongly scintillated source, the modulation index (which measures the typical size of the intensity fluctuations) is unity. 

Now let us imagine that, in addition to a scintillation screen, there is a background lens. That is, the background lens, on its own, results in $N$ distinct images each of which is then modulated by the scintillation screen. We write the voltage and resultant intensity as
\begin{align}
    V_i &= \sum_j^N V^j_i, \\
    I_i &= |\sum_j^N V^j_i|^2,
\end{align}
where, again, each of these $V^j_i$ is a complex, normal variable. Here $i$ indexes the frequency bin, and $j$ indexes the images from the background lens. Since a sum of Gaussians is Gaussian itself, the intensity remains $\chi^2(2)$-distributed, and the modulation index is unity.

Now, however, say the actual intensity is a smoothed version of the underlying intensity due to finite frequency resolution. That is, the observed intensity is
\begin{equation}
    I_k = \Big\langle \Big|\sum_j^N V^j_i\Big|^2 \Big\rangle_{i \in [k + dk, k - dk]},
\end{equation}
where $k$ indexes the re-binned frequency after smoothing. Expanding this gives
\begin{align}
    I_k &= \Big\langle \sum_j^N |V^j_i |^2 + \sum_{j < \ell} V^j_i (V^\ell_i)^*\Big\rangle_{i \in [k + dk, k - dk]} \\
    &= \sum_j^N \big\langle |V^j_i |^2 \big\rangle_{i \in [k + dk, k - dk]},
\end{align}
where the last equality holds if we assume that $V^j_i$ and $V^\ell_i$ are uncorrelated for $j \neq \ell$ (which holds when the scintillation screen resolves the background lens), and we assume the frequency scale of the intensity modulations due to the background lens is finer than the smoothing scale, $dk$. This occurs when the scattering timescale of the background lens is large compared to $dk^{-1}$. 

Now, since each $|V^j_i|^2$ is an independent $\chi^2(2)$ random number, it follows that the expectation value $\overline{I_k} = N K \overline{|V^j_i|^2} = 2 N K$ and ${\rm Var}[I_k] = N K \cdot {\rm Var}[|V^j_i|^2] = 4 N K$, where $K$ is the number of frequency bins we smooth over. From this we obtain
\begin{equation}
    m = \frac{{\rm Var}[I]}{(\overline{I})^2} = \frac{1}{NK}.
\end{equation}
Thus, if the background lens produces $N$ images, the modulation index is suppressed by at least a factor of $N$. The parameter $K$ represents the ratio between the finite frequency resolution of the observation, $\Delta f_{\rm res.}$ and the intrinsic scale of the frequency modulation due to the background lens (assuming the latter is smaller than the former). A lens with scattering timescale $\tau_{\rm scatt.}$ will induce intensity modulations on a frequency scale of $\sim \tau_{\rm scatt.}^{-1}$, from which we obtain: $K \sim \Delta f_{\rm res.} \tau_{\rm scatt.}$. Note that here we have assumed that these lensed images are equally bright, which will generically not be the case. Therefore, in detail, the amount of suppression will depend on not only the number of images but also the relative flux contained in each image. Nevertheless, the basic effect is clear: a background resolving lens suppresses scintillation. The more images the lens produces, or the larger the scattering time scale, the more the modulation index is suppressed.  

\subsection{A more realistic lens model}
\label{sec:realistic}

Here we will introduce a more realistic two-screen lens model that will capture the key effects we are trying to illustrate. Lensing is fully determined by the Fermat potential, or time delay, which for two lenses is given by \cite{Schneider,2020arXiv201003089F}
\begin{align}
    \begin{split}
        T = \frac{1 + z_2}{c} D_{23} \Big[& \frac{1 + z_1}{1 + z_2} \overline{D} \Big(\frac{1}{2} (\boldsymbol{\theta_1} - \boldsymbol{\theta_2})^2 - \psi_1(\boldsymbol{\theta_1}) \Big)\\ 
        & + \frac{1}{2} (\boldsymbol{\theta_2} - \boldsymbol{\theta_3})^2 - \psi_2(\boldsymbol{\theta_2}) \Big],
    \end{split}
\end{align}
where we have define combined distances, $D_{ij} = d_{0i}d_{0j}/d_{ij}$, and $\overline{D} = D_{12} / D_{23}$. The distance $d_{ij}$ is the angular diameter distance between the $i^{\rm th}$ and $j^{\rm th}$ plane, where $j = 0$ refers to the observer plane and $j = 3$ refers to the source plane. The ${\bm \theta}_j$ are the flat-sky angular coordinates for the $j^{\rm th}$ plane and $z_j$ is the redshift of that plane. A plasma lens described by an electron surface over-density $\Sigma_e$ has a lens potential
\begin{equation}
    \psi_j({\bm \theta_j}) =  \frac{c}{D_{j \, j+1} }\frac{\Sigma^j_e({\bm \theta_j}) e^2}{m_e \epsilon_0 \omega^2 (1 + z_j)^2}.
    \label{eq:psi}
\end{equation}
For simplicity, we will consider both lenses to be one-dimensional and oriented in the same direction so that ${\bm \theta_j} \to \theta_j$. For the lenses, we will take a simple sinusoidal phase screen for the nearby lens, $\psi_1 (\theta_1) = -\alpha_1 \sin(\theta_1 / \theta_s)$, and for the second lens we will take a simple rational lens of the form $\psi_2(\theta_2) = -\alpha_2 / (1 + (\theta_2 / \theta_l)^2)$. The parameters $\theta_s$ and $\theta_l$ set the angular scales of the lenses, and the $\alpha$'s set the amplitudes. We have studied rational lenses such as $\psi_2$ extensively elsewhere \citep[see e.g.][]{Jow2021,Jow2023regimes}. We choose a sinusoidal phase screen for the nearby lens in order to model scintillation in the ISM, which has previously been described as being due to corrugated plasma sheets \citep{PenLevin2014,Simard2018}, similar to a sinusoidal sheet. Ultimately, we choose a sinusoidal potential because it allows us to generate many images on the sky, thereby reproducing the Gaussian statistics of strong scintillation, while keeping the number of free parameters small. We will refer to the nearby lens, $\psi_1$, as the ``scintillation screen", and $\psi_2$ as the ``background lens". 

We will choose the angular scale of the background lens, $\theta_l$, to normalize the co-ordinates: $x_1 = \theta_1 / \theta_l$, $x_2 = \theta_2 / \theta_l$, and $y = \theta_3 / \theta_l$. Thus, the time delay for this model can be written as
\begin{align}
    T &= \frac{1 + z_2}{c} \theta^2_l D_{23} \hat{T}, \\
    \hat{T} &= \overline{D} \Big[\frac{1}{2} (x_1 - x_2)^2 - \theta^{-2}_l \alpha_1 \sin(x_1 /\gamma) \Big] + \frac{1}{2}(x_2 - y)^2 - \frac{\theta^{-2}_l \alpha_2}{1 + x_2^2},
\end{align}
where $\gamma = \theta_s / \theta_l$ is the ratio of the charateristic angular scales of the scintillation screen and the lens.

Note that for a plasma lens, the amplitudes of the lens depends on the frequency as $\alpha \sim f^{-2}$ (see Eq.~\ref{eq:psi}). For the sake of simplicity, we will take the amplitudes to be fixed, independent of frequency, as would be the case for gravitational lenses. This makes the numerical calculations easier as one only needs to solve the lens equation once to the find the geometric images, as opposed to finding a new set of images for every frequency. For the purposes of elucidating the effect of multiple images from a background lens on scintillation, it will suffice to assume the images are roughly stationary with frequency. In reality, however, there will always be some frequency above which the lensed images disappear. We include the frequency scalings of the relevant quantities in our discussion of CGM lensing in Section~\ref{sec:CGM}. 

It will now be convenient to re-parameterize in terms of the convergences of the lenses
\begin{align}
    \kappa_1 &\equiv \alpha_1 / \theta^2_s, \\
    \kappa_2 &\equiv \alpha_2 / \theta^2_l,
\end{align}
which are defined such that $\kappa_j = {\rm max} | d^2_{\theta_j} \psi_j(\theta_j)| / 2$. Note that $\alpha_j$ has units of ${\rm rad}^2$, so that $\kappa_j$ is dimensionless. This is the convergence that may be familiar from gravitational lensing. With this re-parametrization, one can write
\begin{align}
    \hat{T} = \overline{D} \Big[\frac{1}{2} (x_1 - x_2)^2 - \gamma^2 \kappa_1 \sin(x_1 /\gamma) \Big] + \frac{1}{2}(x_2 - y)^2 - \frac{\kappa_2}{1 + x_2^2}.
\end{align}
Thus, the dimensionless time delay, $\hat{T}$, depends only on five dimensionless parameters: $\overline{D}$, $\gamma$, $\kappa_1$, $\kappa_2$, and $y$.

In principle, the observed flux is given by an integral over the function $\exp [i\omega T ]$; however, in this work we will be primarily concerned with the refractive effects of plasma lenses in the CGM. That is, we restrict our attention to the geometric optics regime, in which the observed flux is given by a discrete number of images and their magnifications. The image magnifications are encoded in the lens map which determines how points in the lens plane $\bm{x_1}$ are mapped to the source plane $\bm{y}$, which we will call $\bm{y} = \bm{\xi}(\bm{x_1})$. The lens map is determined by the two stationary phase equations: $\nabla_{\bm x_1} T = 0$ and $\nabla_{\bm x_2} T = 0$. It follows that
\begin{align}
\begin{split}
    {\bm y} = {\bm \xi}({\bm x}_1) = &{\bm x_1} + (1 + D)\nabla_{\bm x_1} \psi_1({\bm x_1}) \\
    &+ \nabla_{\bm x_2} \psi_2 \big({\bm x_1} + \nabla_{\bm x_1} \psi_1 ({\bm x_1})\big).
\end{split}
\end{align}
For a given source position $\bm{y}$, we can find locations of the refractive images by inverting the equation $\bm{y} = \bm{\xi}(\bm{x}_1)$. The total observed electric field is a sum over images, given by
\begin{equation}
    F({\bm y}) = \sum_{{\bm x_1} \in {\bm \xi}^{-1}({\bm y})} \frac{1}{\lambda_1^{1/2}({\bm x_1}) \lambda_2^{1/2}({\bm x_1})} e^{i \omega T({\bm x_1}, {\bm x_2}({\bm x_1}), {\bm y})},
\end{equation}
where $\lambda_1$, $\lambda_2$ are the eigenvalues of tha matrix $\nabla_{\bm x_1} {\bm \xi} ({\bm x_1})$. The total observed intensity is $I = |F|^2$, and we each image has a magnification $\mu = |\lambda_1 \lambda_2 |^{-1}$.

\begin{figure*}
    \centering
    \includegraphics[width=2\columnwidth]{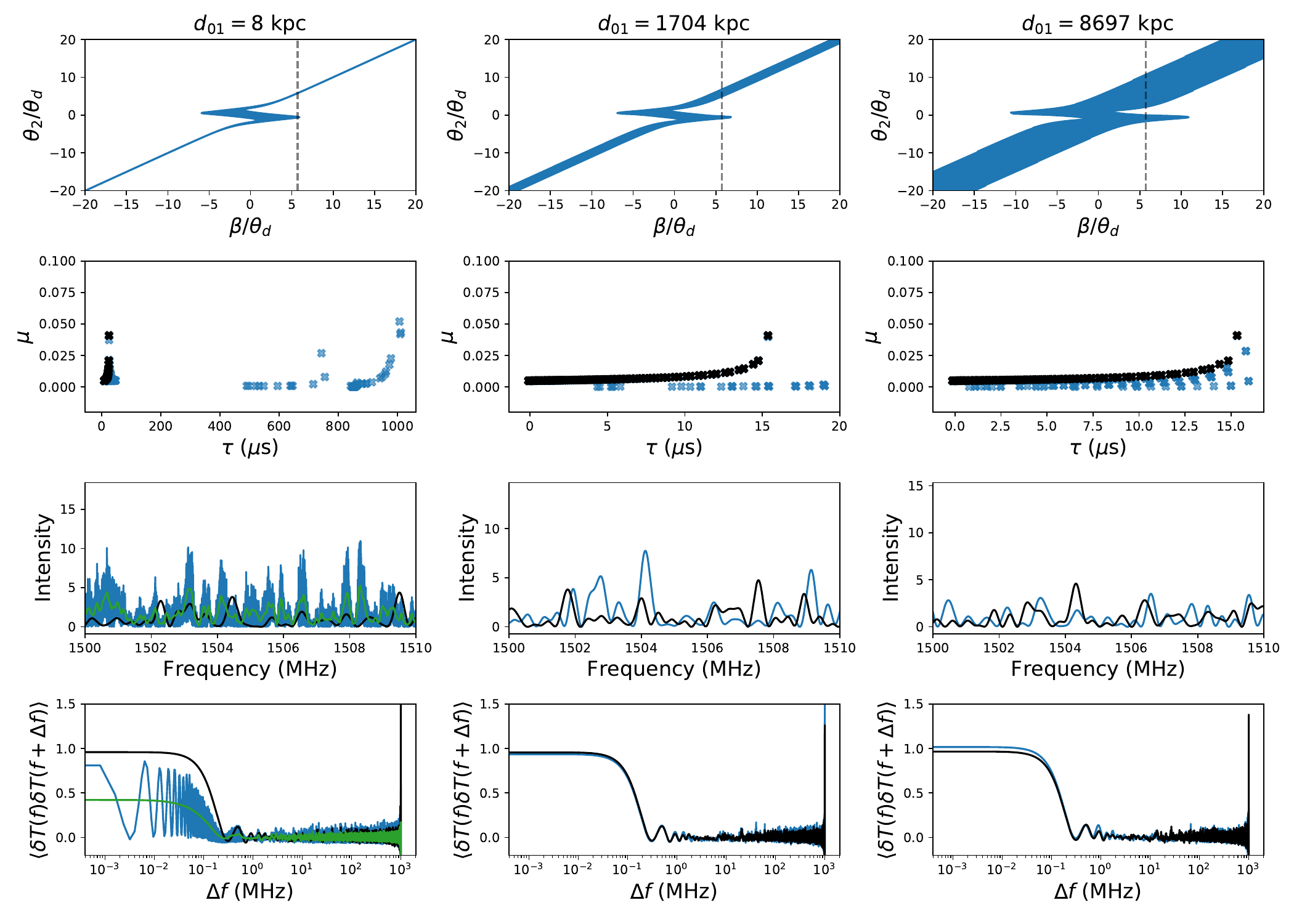}
    \caption{Lensing by the double lens with a sinusoidal phase screen and rational lens with $\kappa_1 = 200$, $\kappa_2 = -10$, $\gamma = 1$, $z_3 = 0.1$, $d_{03} / d_{02} = 2$, $\tau_s = 10\,\mu{\rm s}$, and fixed source position $y = \beta/\theta_l = 5.7$. Columns show the results for different values of $\rho = 0.01, 1, 10$, which corresponds to varying the distance to the sinusoidal screen, $d_{01}$. The top row shows the lens map from the background lens plane to the source plane, $\theta_2 \to \beta$. The grey vertical line corresponds to the source position. The second row shows the magnification and time delay of the images formed by this lens model. The blue dots show the images formed by the background lens and scintillation screen together, whereas the black dots show the images formed by the scintillation screen on its own. The third row shows the intensity modulation as a function of frequency, $I(f)$. Again the blue curves show the full lens model, whereas the black curves show the intensity modulation induced by the scintillation screen on its own. The green curve in the leftmost column shows the result of the blue curve smoothed over some finite frequency resolution, $\delta f = 0.1\,{\rm MHz}$. The bottom row shows the auto-correlation function of these curves, $\langle  \delta T(f) \delta T(f + \Delta f)\rangle$, where $\delta T \equiv (I - \overline{I}) / \overline{I}$. The zero-lag of this auto-correlation function is the modulation index, $m$.}
    \label{fig:lenssim}
\end{figure*}

Now, we would like to determine the typical timescales associated with both the scintillation screen and the lens. Consider the case where $\kappa_1 = 0$, so that the scintillation screen vanishes, and there is only the lens. The typical angular separation between images is given by $\Delta \theta_2 \sim \kappa_2 \theta_l$. It follows that the typical scattering time induced by the lens is
\begin{equation}
    \tau_l = (1 + z_2) \frac{D_{23} \theta^2_l}{2 c} \kappa^2_2.
\end{equation}
Similarly, when $\kappa_2 = 0$, and there is only the scintillation screen, the typical angular separation between images is given by $\Delta \theta_1 \sim \kappa_1 \theta_s$, so that the scintillation timescale is given by
\begin{equation}
    \tau_s = \frac{D_{13} \theta^2_s}{2 c} \kappa^2_1.
\end{equation}
Taking the ratio of these two scattering times, and noting that $D_{13} / D_{23} = \overline{D} / (1 + \overline{D})$, we define
\begin{equation}
    \rho \equiv \frac{\tau_s}{\tau_l} = (1 + z_2)^{-1} \frac{\kappa_1^2}{\kappa_2^2} \frac{\gamma^2 \overline{D}}{1 + \overline{D}}.
\end{equation}
Using this model, we can calculate the effect a background lens has on scintillation. Fig.~\ref{fig:lenssim} shows a summary plot of the lensing computed for three different values of $\rho$: 0.01, 1, 10. The other parameters are kept fixed at $\kappa_1 = 200$, $\kappa_2 = -10$, $\gamma = 1$. A large value of $|\kappa_1|$ is chosen so that the sinusoidal phase screen produces many ($\sim 100$) images in order to mimic scintillation. We choose a value of unity for $\gamma$, as the typical angular scale of a sub-parsec cloud at cosmological distances will be $\theta_l \sim 0.1\,{\rm pc} / 1\,{\rm Gpc} \sim 10\,\mu{\rm as}$, which is the same as the typical angular scale of the scintillated images, $\theta_s \sim 0.1\,{\rm au} / 1\,{\rm kpc} \sim 10\,\mu{\rm as}$. Keeping these parameters fixed and varying $\rho$ causes $\overline{D}$ to change. We choose to keep the source redshift fixed, $z_3 = 0.1$, and the ratio between the source and lens distance fixed, $d_{03} / d_{02} = 2$. Thus, changing $\rho$ is equivalent to changing the distance to the scintillation screen, $d_{01}$. To set the absolute timescale, we choose the scintillation timescale to be fixed at $\tau_s = 10\,\mu{\rm s}$. We choose a fixed source position of $y = \beta / \theta_l = 5.7$. Note that we have re-labled the source position $\theta_3 \to \beta$ to reduce the number of indices floating around and to be consistent with the lensing literature. Note that for this example we chose $\kappa_2 < 0$, which physically corresponds to an under-dense plasma lens. Of course, the expectation is that the CGM cloudlets we hope to model will be over dense. However, our purpose here is to demonstrate the effect of an additional scattering time-scale with as simple an example as possible. The under-dense lens has a simpler phenomenology, with only a single cusp catastrophe. Moreover, a realistic CGM scattering screen will be a projection of all the cool-gas cloudlets transverse to the line-of-sight, and will therefore have over-dense and under-dense regions. 

The top row of Fig.~\ref{fig:lenssim} shows the lens map between the background lens plane and the source plane, $\theta_2 \to \beta$. When $\rho$ is small, the lens map is dominated by the background lens, and one obtains the usual $S$-shape lens map, typical for single-peaked lenses of this kind \citep[see e.g.][]{Clegg1998,2012MNRAS.421L.132P}. As we increase $\rho$, the sinusoidal oscillations from the scintillation screen become more prominent. The second row of Fig.~\ref{fig:lenssim} shows the magnifications, $\mu$, and the relative time delays, $\tau$, between the images produced by the lens. The black dots show the images that form when only the scintillation screen is present, and the blue dots show the images formed by both lenses. One can see that when $\rho$ is small, there are three distinct clusters of images, corresponding to the three images formed by the background lens. Each of these distinct images is independently split into many images by the scintillation screen. Indeed, for $\rho = 0.01$, one can confirm that the scattering timescale associated with the background lens is roughly one-hundred times larger than the scintillation timescale ($\tau_s \sim 10\,\mu{\rm s}$ and $\tau_l \sim 1000\,\mu{\rm s}$). When $\rho$ increases to unity or greater, the scintillation timescale dominates and the images formed by the background lens can no longer be distinguished. The effect of this on the modulation can be seen in the third row of Fig.~\ref{fig:lenssim}. The black curve shows the intensity modulation as a function of frequency, $I(f)$, for the scintillation screen alone. As expected, the black curve is essentially the same across the columns since we have fixed the scintillation timescale to be $\tau_s = 10\,\mu{\rm s}$, which corresponds to a scintillation bandwidth of $\delta f = (2 \pi \tau_s)^{-1} \sim 1\,{\rm MHz}$. The blue curve shows the intensity modulation in the presence of the background lens. When $\rho \geq 1$, the scintillation dominates and the frequency scale of the modulations is again set by the scintillation bandwidth. However, when $\rho < 1$, the frequency scale of the modulations is set by the scattering timescale associated with the background lens. In the case of $\rho = 0.01$, this corresponds to a frequency scale of $0.01\,{\rm MHz}$. The green curve shows what happens when the observation has a finite frequency resolution which is larger than the frequency scale of the intensity modulations. To obtain the green curve, we convolve the blue curve with a Gaussian with a width of $0.1\,{\rm MHz}$, which is a typical value for the frequency resolution of radio observations (see Table~\ref{tab:FRBs}). This leads to intensity modulations which have a similar frequency scale as the un-lensed black curves, but with a visibly reduced modulation index. To see this more clearly, the bottom row of Fig.~\ref{fig:lenssim} shows the auto-correlation function of these different curves, $\langle  \delta T(f) \delta T(f + \Delta f)\rangle$, where $\delta T \equiv (I - \overline{I}) / \overline{I}$. The zero-lag of this function is the modulation index $m$. Essentially, the addition of a new (larger) time scale by the background lens leads to oscillations in the auto-correlation function, which, when averaged over due to finite frequency resolution, results in a suppressed modulation index. In particular, the modulation index is suppressed by about a factor of two. 

So far we have been assuming that the background lens resolves the scintillation screen. That is, a lens that produces geometric images with a physical separation of $\ell$ in the lens plane has an angular resolving power of $\theta_{\rm res.} = \lambda / \ell$. For our background lens with a scattering disk of size $\Delta \theta_2$ at a distance of $d_{02}$, the angular resolving power of the lens is
\begin{equation}
    \theta_{\rm res.} = \frac{\lambda}{d_{02} \Delta \theta_2} \sim \frac{\lambda}{d_{02} \theta_l |\kappa_2|}.
    \label{eq:resolve}
\end{equation}
We have thus far assumed that the background lens resovles the scintillation screen, which is satisfied when the resolution of the background lens is smaller than the spacing between the geometric images of the scintillation screen; i.e., $\theta_{\rm res.} d_{12} < \Delta \theta_1 d_{01}$. Noting that we also assume the scintillation screen is close to the observer such that $d_{12} \approx d_{02}$, this latter condition reduces to $\lambda / \Delta \theta_2 < d_{01} \Delta \theta_1$. An equivalent condition is that the scintillation screen resolves the background lens. It is only when this condition is satisfied that we may use geometric optics (a.k.a. the Eikonal limit) to treat the double-lens system. Essentially, if the scintillation screen does not resolve the images formed by the background lens, then each image is scintillated in the same way, meaning the background lens has no effect on the observed interference pattern. Alternatively, one may say that if the background lens does not resolve the scintillation screen, then the coherence length of the wavefront scattered by the background lens is smaller than the scattering structures in the scintillation screen \citep{Ocker2022}.

\section{Lensing by the CGM}
\label{sec:CGM}

We demonstrated in the previous section that scintillation is suppressed in the presence of a resolving lens. The effect arises simply from the fact that the addition of a second lensing time scale that is larger than the scintillation timescale results in coherent modulation of the intensity on a frequency scale smaller than the scintillation bandwidth. If this new frequency scale is also smaller than the frequency resolution of the observation, the result is that the intensity modulations are smaller than they would be in the absence of the background lens. We demonstrated this effect using a simple rational lens, $\psi_2(x) \propto (1 + x^2)^{-1}$; however, in general, the background lens may be significantly more complicated. The purpose of this paper is to investigate the effect of sub-parsec, cool-gas cloudlets ($T \sim 10^4\,{\rm K}$) in the CGM on FRB scintillation. To that end, we need to describe the refractive scattering properties of these cool gas cloudlets. In particular, we need to determine how many geometric images are formed by these cloudlets and what the typical time-delays between these images are. Following Ref.~\cite{VP2019}, we will assume a model in which the individual cloudlets have a uniform size, $r_c$, and column density, $N^c_e$, with a volume-filling factor throughout the CGM obeying a simple radial power law:
\begin{equation}
    f_V(r) = f^{200}_V \big(\frac{r}{r_{200}} \big)^{-\beta},
    \label{eq:fv}
\end{equation}
where $r_{200}$ is the virial radius and $f^{200}_V$ is the volume-filling factor of the cloudlets at the virial radius. As in Ref.~\cite{VP2019}, we will typically assume $f^{200}_V = 10^{-4}$ unless otherwise stated. Our goal in this paper is to place constraints on this model based on observations of FRB scintillation. 

\subsection{Lenisng properties of CGM cloudlets}
\label{sec:scatteringproperties}

To understand the lensing properties of a halo of cool gas cloudlets, we first want to understand lensing by a single cloud. For a single cloudlet lens of size $r_c$ and electron column density $N^c_e$, the convergence of the lens is given by
\begin{align}
    \begin{split}
        \kappa_c &= \frac{d_{02} d_{23}}{d_{03}} \frac{N^c_e e^2}{m_e r^2_c \epsilon_0 \omega^2 (1 + z_2)^2} \\
        &= 0.3 \,(1+z_2)^{-2} \Big(\frac{d_{02} d_{23} / d_{03}}{1\,{\rm Gpc}} \Big) \\
        &\times \Big( \frac{N^c_e}{10^{17}\,{\rm cm}^{-2}} \Big) \Big( \frac{r_c}{0.1 \, {\rm pc}} \Big)^{-2} \Big( \frac{f}{1\,{\rm GHz}} \Big)^{-2}.
    \end{split}
    \label{eq:kc}
\end{align}
In order for a lens to form multiple geometric images, the convergence needs to be of order unity or greater. Thus, it is already possible for a typical cool-gas cloudlet that forms via the shattering mechanism described in Ref.~\cite{McCourt2018} to lens a background radio source into multiple images. An ensemble of many such cloudlets will produce even more images. 

The typical angular separation between refractive images for a lens of angular size $\theta_l$ and convergence $\kappa_c$ is given by $\theta_l \kappa_c$, so that the scattering time for a single cloudlet lens is given by
\begin{align}
    \begin{split}
        \tau^c_l &= (1 + z_2) \frac{D_{23} \theta^2_l}{2c} \kappa^2_c \\
        &= 35\,\mu{\rm s} \, (1+z_2)^{-3} \Big(\frac{d_{02} d_{23} / d_{03}}{1\,{\rm Gpc}} \Big) \\
        &\times \Big( \frac{N^c_e}{10^{17}\,{\rm cm}^{-2}} \Big)^2 \Big( \frac{r_c}{0.1 \, {\rm pc}} \Big)^{-2} \Big( \frac{f}{1\,{\rm GHz}} \Big)^{-4}.
    \end{split}
\end{align}
Typical scintillation timescales are $\sim 10\,\mu{s}$, therefore even lensing by a single cloudlet may lead to a suppression of scintillation following the effect we describe in Section~\ref{sec:lensing}. In general, any given sightline through a CGM halo will intersect many such cloudlets. Assuming a volume filling factor of the form Eq.~\ref{eq:fv}, the average number of cloudlets that a sight-line through the CGM with impact parameter $b$ intersects is given by
\begin{align}
    \begin{split}
        N(b) &=  \int_{-r_{200}}^{r_{200}} \frac{\pi r^2_c}{\frac{4}{3} \pi r^3_c } f_V(\sqrt{\ell^2 + b^2}) d\ell \\
        &\sim f^{200}_V r_{200} r_c^{-1},
    \end{split}
    \label{eq:Nb}
\end{align}
where the latter scaling relation is obtained within the virial radius ($b < r_{200}$) and $\beta \lesssim 1$.

For an ensemble of cloudlets, projected onto the lens plane, the total column density scales with the number of intersected cloudlets, $N_e \sim N(b) N^c_e$. However, since for lensing what matters is the \textit{relative} electron surface density, the typical convergence of a column of cloudlets will scale with the rms fluctuations of the column density, i.e. $\kappa \sim \kappa_c \sqrt{N}$. Thus, we have the total scattering time from an ensemble of cloudlets is $\tau_l \sim N \tau^c_l$, which is
\begin{align}
\begin{split}
    \tau_l \sim 10.5\,{\rm ms} \, &(1 + z_2)^{-3} \Big( \frac{f^{200}_V}{10^{-4}} \Big) \Big( \frac{r_{200}}{100\,{\rm kpc}} \Big) \Big(\frac{d_{02} d_{23} / d_{03}}{1\,{\rm Gpc}} \Big) \\
    &\Big( \frac{N^c_e}{10^{17}\,{\rm cm}^{-2}} \Big)^2 \Big( \frac{r_c}{0.1 \, {\rm pc}} \Big)^{-3} \Big( \frac{f}{1\,{\rm GHz}} \Big)^{-4}.
\end{split}
\label{eq:taud}
\end{align}
This refractive scattering time is only sensible when the lens in question produces multiple images (i.e. there needs to be more than one image in order for the time delay between images to be well defined). Thus, we would like to know how many images an ensemble of cloudlets is likely to produce. For an ensemble of one-dimensional lenses, each with convergence $\kappa$, spaced uniformly along a line with a spacing given by $\theta_l$, the number of additional images formed scales like $\kappa$, since $\kappa$ sets the maximum angular separation, $\beta / \theta_l$, from the lens the source can attain such that the lens forms more than one image. For an ensemble of two-dimensional lenses, such as a ensemble of cool-gas cloudlets projected onto the plane of the sky, the number of additional images will scale like $\kappa^2$. Thus, the number of images formed by a halo of CGM cloudlets, $N_{\rm im.}$, will scale like $N_\mathrm{im.} - 1 \sim N \kappa_c^2$, where again $N$ is the typical number of cloudlets intersected by a given sight-line. It follows that
\begin{align}
\begin{split}
    N_\mathrm{im.} - 1 \sim 20 \, & (1+z_2)^{-4} \Big( \frac{f^{200}_V}{10^{-4}} \Big) \Big( \frac{r_{200}}{100\,{\rm kpc}} \Big) \Big(\frac{d_{02} d_{23} / d_{03}}{1\,{\rm Gpc}} \Big)^2 
    \\
    &  \Big( \frac{N^c_e}{10^{17}\,{\rm cm}^{-2}} \Big)^2 \Big( \frac{r_c}{0.1 \, {\rm pc}} \Big)^{-5} \Big( \frac{f}{1\,{\rm GHz}} \Big)^{-4}.
\end{split}
\label{eq:Nim}
\end{align}
Thus, for typical parameters of the CGM cloudlets, an FRB with a sight-line passing within $r_{200}$ of a galaxy will be lensed into roughly twenty refractive images, with a lensing time-scale of $10.5\,{\rm ms}$. This would lead to intensity modulation with a typical frequency scale of $10^{-3}\,{\rm MHz}$, which is well bellow typical frequency resolutions. Thus, if the observed cool gas in the CGM were to exist primarily in cloudlets of this form, we would expect FRB scintillation to be strongly suppressed. Following our calculation in Section~\ref{sec:simple}, the modulation index will be suppressed by roughly $N_{\rm im} \Delta f_{\rm res.} \tau_l \sim 10^4$ for a frequency resolution of $0.1\,{\rm MHz}$.

So far, we have computed the refractive scattering properties of the CGM cloudlets using simple scaling arguments. We can, however, verify these results numerically by simulating the lens potential that an FRB with impact parameter $b$ relative to the galactic centre of some galaxy would see. For a given impact parameter $b$, we divide the sight-line from $-r_{200}$ to $+r_{200}$ into 40 boxes of dimensions $100 r_c \times 100 r_c$ transverse to the line-of-sight, and a height of $0.05 r_{200}$ along the line-of-sight. For each box, we uniformly distribute $n$ cloudlets, where $n$ is a Poisson random variable with mean given by $f_V(r) V_{\rm box} / V_c$, where $r$ is the co-ordinate for the centre of the box, $V_{\rm box} = 100 r_c \times 100 r_c \times 0.05 r_{200}$, and $V_c = \frac{4}{3} \pi r^3_c$. In doing so, we ensure that the volume filling density of the randomly generated cloudlets obeys Eq.~\ref{eq:fv}. To determine the lens potential, we then project the generated ensemble of clouds onto the plane of the sky, assuming that each cloudlet has a Gaussian density profile of $N_e({\bm r}) = N^c_e e^{-2 ({\bm r} - {\bm r}_i)^2 / r^2_c}$, where ${\bm r}_i$ is the co-ordinate of the centre of the cloud. 

\begin{figure}
    \centering
    \includegraphics[width=0.6\columnwidth]{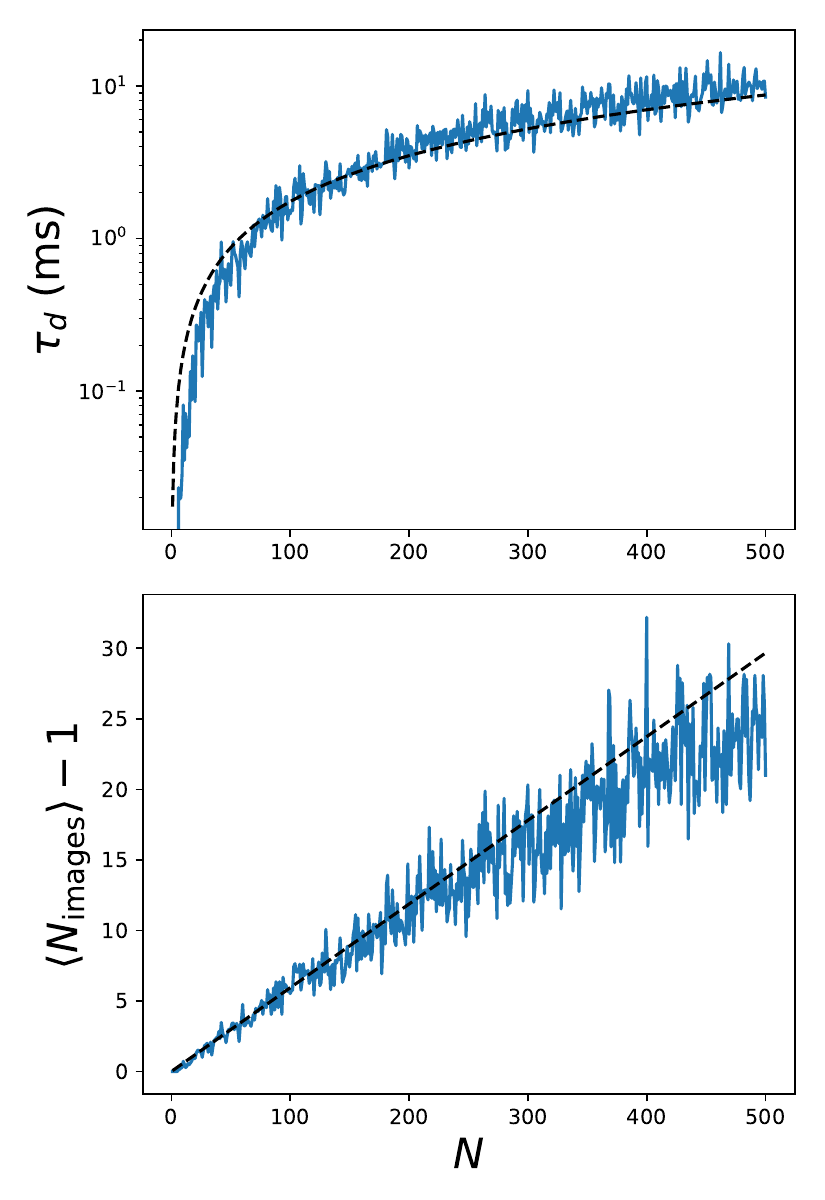}
    \caption{Average scattering time (top) and number of refractive images (bottom) formed by an ensemble of cool gas cloudlets as a function of the average number of cloudlets intersected by a given sightline (Eq.\ref{eq:Nb}). We take the cloudlets to have uniform size, $r_c = 0.1\,{\rm pc}$, with a uniform maximum column density, $N^c_e = 10^{17}\,{\rm cm}^{-2}$, and a symmetric Gaussian density profile. We take the galaxy to have a virial radius of $r_{200} = 100\,{\rm kpc}$ and the clouds to have a power-law volume filling factor given by Eq.~\ref{eq:fv}, with $\beta = 0.2$. We vary the number of intersected cloudlets by varying the overall amplitude of the volume filling factor, $f^{200}_V$. We choose a distance parameter of $d_{02}d_{23}/d_{03} = 1\,{\rm Gpc}$ for the lensing geometry, an observing frequency of $1\,{\rm GHz}$, and an impact parameter of $b = r_{200}$. }
    \label{fig:Nscaling}
\end{figure}

Once we have generated the projected electron density and thereby the lens potential via Eq.~\ref{eq:psi}, we can compute the number of images formed by the lens and the associated time delay. Fig.~\ref{fig:Nscaling} shows the result of this calculation for a galaxy with virial radius $r_{200} = 100\,{\rm kpc}$, $\beta = 0.2$, and an impact parameter of $b = r_{200}$. We choose the cloudlet sizes to be $r_c = 0.1\,{\rm pc}$, the observing frequency to be $f = 1\,{\rm GHz}$, and the distance parameter to be $d_{02} d_{03} / d_{23} = 1\,{\rm Gpc}$. We compute the number of images and mean time delay between images for multiple sightlines in a range of $[b - 100 r_c, b + 100 r_c]$ and report the average of these values in Fig.~\ref{fig:Nscaling}. In order to vary the number of clouds the sightline intersects, we vary $f_V^{200}$, and computing $N(b)$ using Eq.~\ref{eq:Nb}, we plot the scattering time and number of images as a function of the number of clouds intersected. In doing so, we can confirm the scalings: $\tau_l \sim N \kappa_c$ and $N_{\rm im} - 1 \sim N \kappa$. The blue curves in Fig.~\ref{fig:Nscaling} show the numerically computed values and the dashed lines show the analytic approximations in Eqs.~\ref{eq:taud} and \ref{eq:Nim}.

\subsection{Constraints on CGM cloudlets}
\label{sec:constraints}

Thus far, we have demonstrated that when a lens produces multiple refractive images that are optically resolved by the scintillation screen, the observed modulation index of scintillation is suppressed when the scattering timescale of the lens is larger than $\sim\Delta f_{\rm res.}^{-1}$, where $\Delta f_{\rm res.}$ is the frequency resolution of the observations. In particular, the modulation index is expected to be suppressed by a factor of $\sim N_{\rm im} \Delta f_{\rm res.} \tau_l$, where $N_{\rm im}$ is the number of images produced by the lens and $\tau_l$ is the scattering timescale associated with the lens. In the previous section, we showed that spherical CGM cloudlets of cool, ionized gas with a characteristic radius of $r_c = 0.1\,{\rm pc}$ and electron column density of $N_e = 10^{17}\,{\rm cm}^{-2}$ (consistent with the shattering picture proposed by Ref.~\cite{McCourt2018}) will generically result in refractive scattering of FRBs, yielding $N_{\rm im} \sim 20$ refractive images with a typical scattering timescale of $\tau_l \sim 10\,{\rm ms}$. Given that typical frequency resolutions of FRB observations are on the order of $\Delta f_{\rm res.} = 0.1\,{\rm MHz}$, or $\Delta f_{\rm res.}^{-1} = 10\,\mu{\rm s}$, then we can infer that if the CGM is ubiquitously populated by sub-parsec cloudlets, then any FRB with a sight-line that passes within the virial radius of a galactic halo along its line of sight will not scintillate, or will do so very weakly. 

\begin{figure}
    \centering
    \includegraphics[width=\columnwidth]{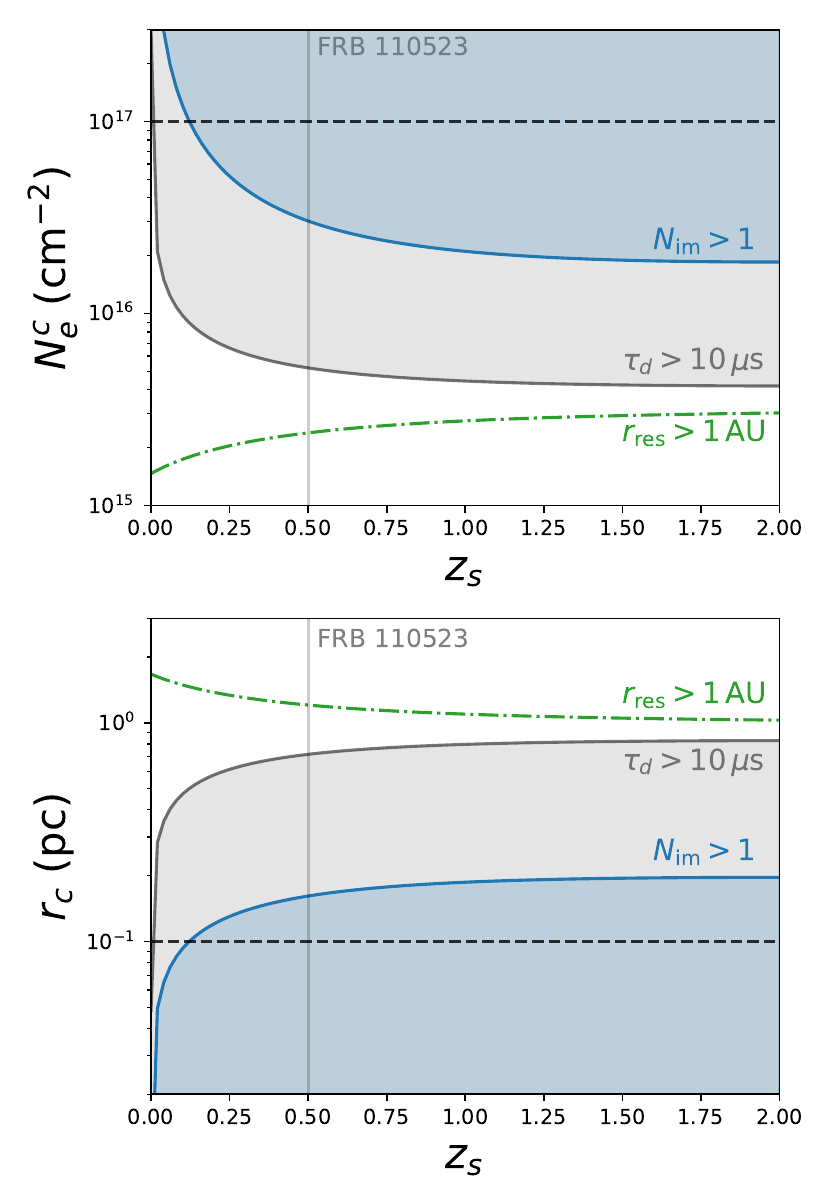}
    \caption{Conditions under which FRB scintillation is expected to be suppressed for an FRB at a given source redshift, $z_s$, with a sightline passing within the virial radius of galactic halo. We assume the galactic halo comprises an ensemble of cool-gas cloudlets of uniform radius $r_c$ and electron column density $N_e$, with a volume filling factor of $f^{200}_V = 10^{-4}$ at the virial radius. Scintillation is suppressed when the cloudlets produce many refractive images, $N_{\rm im} > 1$, which produce frequency modulations on a scale finer than the frequency resolution of the observation, which we take to be $\Delta f_{\rm res.} = 0.1\,{\rm MHz}$. This latter condition corresponds to a scattering timescale of $\tau_d > 10\,\mu{\rm s}$. Inverting Eqs.~\ref{eq:taud} and \ref{eq:Nim}, we compute the values of $N_e$ for which scintillation is suppressed as a function of source redshift for fixed $r_c = 0.1\,{\rm pc}$ (top panel), as well as the values of $r_c$ for which scintillation is suppressed for fixed $N_e = 10^{17}\,{\rm cm}^{-2}$. The shaded regions show the excluded values of these parameters given a positive detection of scintillation. The green curves show the condition that CGM lens resolves the scintillation screen; however, note that whenever $N_{\rm im} > 1$, this condition is always satisfied. The horizontal dashed line shows the fiducial values of the parameters $r_c$ and $N_e$ as predicted by the shattering picture proposed by Ref.~\cite{McCourt2018}. The vertical grey line shows the inferred redshift of FRB\,110523, which was the first FRB to be confirmed to scintillate due to multi-path propagation in the Milky Way ISM, with a modulation index of $m = 0.27$.}
    \label{fig:constraints}
\end{figure}

Fig.~\ref{fig:constraints} shows potential constraints that may be placed on the physical parameters of cool-gas, CGM cloudlets based on the preceding discussion. Using Eqs.~\ref{eq:taud} and \ref{eq:Nim}, we can place constraints on the physical size and column density of the cloudlets given an observation of scintillating FRB at a given redshift, $z_s$, whose sight-line passes within the virial radius of an intervening galactic halo. That is, if the inferred scattering time is greater than $\Delta f_{\rm res.}^{-1} \sim 10\,\mu{\rm s}$, and the number of refractive images is $N_{\rm im} > 1$, then scintillation will be suppressed for that FRB. The top panel of Fig.~\ref{fig:constraints} shows the constraints on the electron column density of the CGM cloudlets that could be derived from a scintillating FRB as a function of source redshift: column densities above the grey and blue curve yield refractive images with scattering times greater than $10\,\mu{\rm s}$ and $N_{\rm im} > 1$, respectively. Thus, column densities above those two curves for a given redshift would be ruled out by a scintillating FRB at that redshift. As we have discussed, another criterion that must be met is that the scintillation screen in the ISM be resolved by the refractive images (or vice versa). Since pulsar scintillation observations reveal typical scattering length scales of $\sim 1\,{\rm AU}$ \citep[see e.g.][]{Brisken2010}, then using Eq.~\ref{eq:resolve}, we also require $d_{01} \theta_{\rm res.} > 1\,{\rm AU}$, shown by the green curve in Fig.~\ref{fig:constraints}. Note, however, that generically whenever $N_{\rm im} > 1$, the scintillation screen is always resolved. The bottom panel of Fig.~\ref{fig:constraints} likewise shows constraints one might obtain on the characteristic radius of the cloudltes, $r_c$, for fixed column density $N_e$. Of course, in general, the physical mechanism that forms and sustains these cloudlets will cause the radius and column density of the cloudlets to be related to each other. Nevertheless, for the values of $r_c = 0.1\,{\rm pc}$ and $N_e = 10^{17}\,{\rm cm}^{-2}$ predicted by Ref.~\cite{McCourt2018}, the detection of a scintillating FRB whose sightline passes through a galactic halo at most source redshifts would disfavour the spherical cloudlet model. We note, however, that these nominal values are close to marginally disfavoured.

We argue that current FRB scintillation observations -- while not nearly comprehensive enough to place precise constraints on the properties of cool-gas cloudlets in the CGM -- already disfavour the ubiquitous presence of such cloudlets. In particular, Ref.~\cite{VP2019} show that most FRB sightlines will pass within the virial radius of at least one galactic halo (and, in some cases, many times more). Thus, if these sub-parsec cloudlets of cool gas were ubiquitous, then our calculations suggest that FRBs should rarely scintillate, and should do so only very weakly. Yet, FRB scintillation does not seem to be a rare, or weak phenomenon. For example, Ref.~\cite{Schoen2021} find that within a sample of twelve FRBs, ten have observable scintillation. Ref.~\cite{Sammons2023} find that of a sample of nine FRBs, three show positive evidence for scintillation, while two show ambiguous evidence for scintillation. While these are by no means unbiased samples, and therefore one must be careful in drawing quantitative conclusions about the relative frequency of FRB scintillation, nevertheless, it is clear that FRB scintillation is not a rare phenomenon. Table~\ref{tab:FRBs} shows a list of scintillating FRBs with published modulation indices. Two of the FRBs listed, FRBs 20220912A \citep{WuMain2023} and 20201124A \citep{Main2023}, have modulation indices consistent with unity, and are thus, strongly scintillated, highly disfavouring any amount of scattering along the line of sight at the observing frequencies. 

Another case of particular relevance to this discussion is FRB\,20190520B, which has been confirmed to scintillate, although the low signal-to-noise of the burst makes it challenging to measure a precise modulation index \citep{Ocker2022}. Recently, the sight-line of this FRB was found to pass within the virial radius of at least two foreground galaxy clusters, and, as a result, the FRB has a particularly high dispersion measure (DM) given its redshift \citep{2023arXiv230605403L}. The fact that this FRB both scintillates and is known to have excess DM due to passing through multiple CGM halos presents a challenge to the existence of sub-parsec, cool-gas cloudlets in the CGM. Therefore, we argue that the present observational status of FRB scintillation does not support the ubiquitous presence of such  cloudlets. More detailed analyses of a larger sample of scintillating FRBs will be needed to make more robust conclusions. 

\begin{table*}
    \centering
    \begin{tabular}{ c | c | c | c | c | c | c | c | c}
        FRB & Publication & Type  & DM (pm cm$^{-3}$) &  Inferred redshift & f (MHz) & $\Delta f_{\rm res.}$ (MHz) & $\Delta f_s$ (MHz) & $m$ \\ \hline
        110523 & Ref.~\cite{Masui2015} & Single burst & 623.30 & <0.5 & 700 - 900 &0.05 & 1.24 & 0.27 \\ \hline
        121101 & Ref.~\cite{Hessels2019} & Repeater & 560.57 & 0.2 & 1100 - 1700 & 0.0039 & 0.058 & 0.25 \\ \hline
        180916.J0158+65c & Ref.~\cite{Marcote2020} & Repeater & 348.76 & 0.03 & 1630 - 1760 & 0.0156 & 0.059 & 0.06 \\ \hline
        20220912A & Ref.~\cite{WuMain2023} & Repeater & 220 & 0.077 & 1100 - 1500 &0.1 & 0.39 & 1\\ \hline
        20190608B & Ref.~\cite{Sammons2023} & Single burst & 338.7 & 0.118 & 1200 - 1400 & 0.1 & 1.4 & 0.78 \\ \hline
        20210320C & Ref.~\cite{Sammons2023}  & Single burst & 384.8 & - & 700 - 950 & 0.1 & 0.91 & 0.68 \\ \hline
        20201124A & Ref.~\cite{Sammons2023}  & Repeater & 410.83 & 0.098 & 695 - 735 & 0.01 & 0.14 & 0.35 \\ \hline
        20201124A & Ref.~\cite{Main2023} & Repeater & 410.83 & 0.098 & 1270 - 1470 & 0.195 & 1.34 & 1 \\ \hline
    \end{tabular}
    \caption{Scintillating FRBs with published modulation indices. The inferred redshifts for 20190608B and 20201124A are from Ref.~\cite{2022ApJ...931...88C}. The quantity $\Delta f_s$ refers to the measured scintillation bandwidth. The frequency range, $f$ over which the observations were made, as well as the frequency resolution, $\Delta f_{\rm res.}$, are reported.}
    \label{tab:FRBs}
\end{table*}

\section{Discussion}
\label{sec:discussion}

We have argued that the existence of a mist of sub-parsec, cool-gas cloudlets throughout the CGM of galaxies is broadly inconsistent with observations of FRB scintillation. Nevertheless, we know from absorption studies that cool gas with a low total volume, but large covering factor is common in the CGM of galaxies, suggesting that the cool gas is not diffuse throughout the CGM, but exists in many isolated clumps. How do we reconcile these two observations? One straightforward possibility is that the cool gas does not form roughly spherical cloudlets, but rather highly anisotropic structures, such as sheets or filaments. Both cool-gas sheets and filaments are equally as plausible as spherical cloudlets as an explanation for the low total volume but high covering factor observed in absorption measurements, as these absorption observations cannot distinguish between different geometries of cool-gas structures. However, as has been shown, cool-gas confined to sub-parsec cloudlets will yield strong lensing phenomena that have not been observed. By relaxing the confinement of the gas along one or two dimensions (i.e. by allowing the gas to form filaments or sheets) alleviates this tension, as, for example, a sheet of ionized plasma will only lead to multi-path propagation of an FRB when the inclination angle of the sheet is close to aligned with the line-of-sight of the FRB. Filaments are a less promising solution to the problem, as filaments still lead to large transverse phase gradients in the scattered wavefront in at least one direction regardless of the orientation of the filament. Thus, we propose that the commonly observed cool ($T \sim 10^4\,{\rm K}$) gas in the CGM forms sheet-like structures, as opposed to cloudlets. We do not, however, present here a plausible physical picture for how these sheets might form and persist in the CGM; we simply point out that this geometry may resolve the present observational conundrum.   

Extrapolating from the few FRBs with published modulation indices (see Table~\ref{tab:FRBs}), many FRBs appear to scintillate with modulation indices less than unity by a factor of two or three. This is in contrast with scintillating pulsars. In recent decades, it has been discovered that a large fraction of pulsars scintillate due to scattering dominated by a single thin, highly anisotropic plasma sheet in the ISM along the pulsar's line of sight \citep{2001ApJ...549L..97S, 2004MNRAS.354...43W, Brisken2010}. These scintillating sheets generally induce intensity modulations in pulsars in both the frequency and time domain with modulation indices close to unity \citep[see e.g.][for a survey of scintillating pulsars]{Stinebring2022}. A detailed analysis of the scintillation of the Vela pulsar revealed a modulation index that was within $0.01 \%$ of unity \citep{Johnson2012}. Some pulsars are known to scintillate only very weakly ($m \ll 1$). In the case of the Crab pulsar ($m < 0.02$), the modulation index is known to be highly suppressed due to a secondary scattering screen in the Crab nebula which is resolved by the scintillation screen in the ISM (the same effect we have been describing for FRBs) \citep{Main2021}. However, modulation indices with only a modest suppression, as observed in FRBs, are less frequently seen in pulsars; that is, pulsars are seem to scintillate either strongly, or very weakly. If the same structures in the ISM are responsible for both pulsar and FRB scintillation, as they presumably are, then this discrepancy in the relative strength of scintillation between the two source populations is unexpected. 

If the CGM comprises a network of sheets of cool gas with sub-parsec thickness, then this may already provide an explanation for the fact that FRBs appear to scintillate more weakly than pulsars. We have shown that the suppression of the modulation index scales as the number of refractive images, $\sim N_{\rm im}$. While lensing by a single CGM halo comprising an ensemble of spherical cloudlets with sub-parsec radii may result in many tens of refractive images (thereby effectively washing out any scintillation), an ensemble of \textit{sheets} with sub-parsec thickness may produce only a small number of refractive images, as only a small number of the sheets will ever contribute to the lensing. This follows from the fact that only sheets closely aligned with the line-of-sight of the FRB will induce phase gradients large enough to lens the FRB at all. Moreover, while absorption observations suggest that such sheets must have sub-parsec thickness, the transverse inhomogeneities of the sheets may have an entirely different characteristic scale, leading to smaller scattering times. Thus, cool-gas sheets in the CGM may well lead to only modest suppression of FRB scintillation, consistent with the some of the FRB modulation indices published so far. A specific case of interest is FRB 20201124A, which was found to have a modulation index consistent with unity in the $1270 - 1470\,{\rm MHz}$ band \citep{Main2023}, and a modulation index of $m = 0.35$ in the $695 - 735\,{\rm MHz}$ band \citep{Sammons2023}. Since plasma lenses increase in strength as $\kappa \sim f^{-2}$, a relative suppression of the modulation index at lower frequencies may indicate the presence of scattering. Further analysis of this burst would be needed to confirm this picture. If plasma sheets are present in the CGM, they would likely need to be sustained by dynamically relevant magnetic fields, providing further motivation for FRB polarimetric studies to probe the magnetic properties of the CGM. 

While sheet-like geomteries may explain the current scintillation observations, we note that our arguments herein only disfavour the nominal values of the cloudlet model considered by Refs.~\cite{McCourt2018} and \cite{VP2019}. As can be seen in Fig.~\ref{fig:constraints}, if the characteristic radius of the cloudlets exceeds the nominal $0.1\,{\rm pc}$ value, then the ensemble of cloudlets will not produce enough refractive images to suppress scintillation. Similarly, a spectrum of cloudlet sizes may be sufficient to reduce the lensing effect so that only some fraction of CGM halos produce enough refractive scattering to suppress scintillation. That said, we point out that in our lensing simulations we have only considered smooth cloudlets of a charactersitic size with no subtructure. In the absence of an unambiguous prediction of the cloudlet substructure, we have chosen to focus on the guaranteed refractive effect from the density fluctuations at the outer scale. But as Ref.~\cite{VP2019} show, turbulent substructure down to the diffractive scale ($\sim 10^{11}\,{\rm cm})$ is already ruled out. While we have considered the minimal case with no substructure here, we stress that substructure of any kind will only enhance the lensing effect and subsequent suppression of scintillation. Whatever the case, Milky Way ISM scintillation observations have considerable potential to probe plasma lensing in the CGM.

Whether or not lensing at all explains the observed modulation indices of FRBs is also speculative. There are other possibilities for why FRB modulation indices are modestly suppressed which do not invoke any kind of lensing; for example, nano-structure in the intrinsic emission profile can suppress intensity correlations in scintillation \citep{2023ApJ...945..115L}. However, nano-structure can at most suppress scintillation by a factor of $1/\sqrt{3}$, and in the case of the repeater FRB 20201124A, the spectra of nearby bursts are found to correlate at $\sim 100\%$, suggesting that there is no intrinsic fine spectral structure \citep{Main2023}. We hope, therefore, that this work provides additional impetus for the detailed study of FRB scintillation as a potential probe of CGM physics and lensing in general.

\section{Conclusion}
\label{sec:conclusion}

Scintillation of FRBs is caused by scattering in plasma screens in the Milky Way ISM. When a refractive lens along the line of sight forms multiple images that resolve the scintillation screen, the amount of scintillation observed, as measured by the modulation index, $m$, may be suppressed. This suppression occurs when the scattering time associated with the lensed images leads to intensity modulations on a frequency scale that is smaller than the frequency resolution of the observation. Essentially, the intensity modulations induced by the lens are smoothed over due to the finite resolution, leading to a smaller modulation index than in the absence of the lens. In this paper, we have shown that sub-parsec, cool-gas cloudlets in the CGM of the type proposed by Ref.~\cite{McCourt2018} will generically lens FRBs in such a way as to suppress scintillation by the Milky Way ISM. Therefore, since most sight-lines are expected to pass through at least one CGM halo, and since FRB scintillation is neither a rare nor weak phenomenon, we argue that the current observations disfavour the existence of these cloudlets. We suggest a sheet-like geometry for the cool gas in the CGM as a way to accommodate both absorption studies, which demonstrate the existence of a cool-gas phase in the CGM with a large covering fraction, and the lack of lensing signals from this gas.

\acknow{We receive support from Ontario Research Fund—research Excellence Program (ORF-RE), Natural Sciences and Engineering Research Council of Canada (NSERC) [funding reference number RGPIN-2019-067, CRD 523638-18, 555585-20], Canadian Institute for Advanced Research (CIFAR), Canadian Foundation for Innovation (CFI), the National Science Foundation of China (Grants No. 11929301),  Thoth Technology Inc, Alexander von Humboldt Foundation, and the Ministry of Science and Technology(MOST) of Taiwan(110-2112-M-001-071-MY3). Computations were performed on the SOSCIP Consortium’s [Blue Gene/Q, Cloud Data Analytics, Agile and/or Large Memory System] computing platform(s). SOSCIP is funded by the Federal Economic Development Agency of Southern Ontario, the Province of Ontario, IBM Canada Ltd., Ontario Centres of Excellence, Mitacs and 15 Ontario academic member institutions.
 Cette recherche a \'{e}t\'{e} financ\'{e}e par le Conseil de recherches
en sciences naturelles et en g\'{e}nie du Canada (CRSNG), [num\'{e}ro de
r\'{e}f\'{e}rence 523638-18,555585-20 RGPIN-2019-067]. }

\showacknow{} 

\bibsplit[20]

\bibliography{biblio}

\end{document}